\documentclass[12pt]{iopart}

\usepackage{graphicx,color}

\begin{document}

\title[Structural characterisation of polycrystalline colloidal monolayers]{Structural characterisation of polycrystalline colloidal monolayers in the presence of aspherical impurities}

\author{Andrew T Gray$^1$, Elizabeth Mould$^1$, C Patrick Royall$^{1,2,3}$ and Ian Williams$^{1,2,3}$}
\address{$^1$ School of Chemistry, Cantock's Close, University of Bristol, BS8 1TS, UK}
\address{$^2$ H.H. Wills Physics Laboratory, Tyndall Ave., Bristol, BS8 1TL, UK}
\address{$^3$ Centre for Nanoscience and Quantum Information, Tyndall Avenue, Bristol BS8 1FD, UK}
\ead{ian.williams@bristol.ac.uk}
\date{\today}

\begin{abstract}
\textbf{
Impurities in crystalline materials introduce disorder into an otherwise ordered structure due to the formation of lattice defects and grain boundaries. The properties of the resulting polycrystal can differ remarkably from those of the ideal single crystal. Here we investigate a quasi-two-dimensional system of colloidal spheres containing a small fraction of aspherical impurities and characterise the resulting polycrystalline monolayer. We find that, in the vicinity of an impurity, the underlying hexagonal lattice is deformed due to a preference for 5-fold co-ordinated particles adjacent to impurities. This results in a reduction in local hexagonal ordering around an impurity. Increasing the concentration of impurities leads to an increase in the number of these defects and consequently a reduction in system-wide hexagonal ordering and a corresponding increase in entropy as measured from the distribution of Voronoi cell areas. Furthermore, through both considering orientational correlations and directly identifying crystalline domains we observe a decrease in the average polycrystalline grain size on increasing the concentration of impurities. Our data show that, for the concentrations considered, local structural modifications due to the presence of impurities are independent of their concentration, while structure on longer lengthscales (\textit{i.e} the size of polycrystalline grains) is determined by the impurity concentration.
}
\end{abstract}

\maketitle

\section{Introduction}
In theory, the constituent particles of a crystalline material are perfectly ordered on a periodic lattice and thus exhibit long-ranged translational and orientational order. In two dimensions, the crystalline structure adopted by a system of identical circular particles interacting isotropically in the absence of external fields is invariably hexagonal. However, at finite temperature, long wavelength fluctuations that would have divergent energy cost in three dimensions are stable in 2d, reducing translational order \cite{peierls1935,mermin1966,mermin1968,gasser2010}. The result is that, except in the close packed limit, two-dimensional systems exhibit only quasi-long-ranged translational order characterised by algebraically decaying translational correlation functions. In spite of this, orientational ordering remains long-ranged in the two-dimensional hexagonal crystal \cite{gasser2010,murray1987,bernard2011}. Intermediate between the hexagonal crystal and the isotropic fluid is the hexatic phase characterised by exponentially decaying translational correlations (short ranged translational ordering) and algebraically decaying orientational correlations (quasi-long-ranged orientational ordering) \cite{gasser2010,murray1987,bernard2011,halperin1978}. 

It is often useful to consider ideal single crystals consisting of an infinite array of particles located on a lattice. The experimental reality, however, is rarely so perfect. Real crystalline materials commonly feature grain boundaries and other defects which reduce their long-ranged order and modify their properties compared to a single crystal. For example, the yield stress of a polycrystalline material depends upon the size of crystalline grains. In the micrometre size regime, smaller grains result in a stronger material \cite{hall1951,yip1998} while nanocrystals show the opposite behaviour: softening on decreasing grain size \cite{yip1998,chokshi1989,schiotz1998}. The crossover between these behaviours is material dependent but is typically located at a lengthscale in the tens of nanometres. Similarly, the optical properties of crystalline materials are also dependent on the distribution of defects due to, for example, scattering from grain boundaries \cite{sotelo2003} or pores \cite{peelen1974}. Defects and grain boundaries may arise from the presence of impurities in a crystal \cite{devilleneuve2009,dullens2008,millett2006} and thus controlling the concentration and morphology of impurities allows the tuning of material properties.

Here we investigate a quasi-two-dimensional colloidal crystal in the presence of aspherical impurities and demonstrate their effect on the structural properties of the system. Colloids are an attractive model system for the study of many general phenomena due to their experimentally accessible time and length scales allowing real-space observation and the extraction of particle-level information. Two-dimensional colloidal crystals are of interest for a range of applications from templating the growth of three-dimensional crystals (colloidal epitaxy) \cite{vanblaaderen1997,allard2004}, through designing phononic \cite{baumgartl2007} and photonic \cite{allard2004,kumacheva2002} bandgap materials, to surface coatings \cite{zhao2005}, adhesives and lubrication technologies as well as offering fundamental insight into the behaviour of atomic and molecular materials \cite{poon2004}.

Previous investigation has focused on the effect of impurities that are either much larger \cite{devilleneuve2009,dullens2008,engelbrecht2010} or much smaller \cite{ghofraniha2012} than their host colloids on crystallisation in three dimensions. These studies demonstrate that the texture of a colloidal polycrystal can be controlled through suitable choice of impurity size and concentration. 

Here we focus on the intermediate regime where colloid and impurity are of similar size and disorder results from the aspherical shape of the impurity particles. In Section \ref{secMethods} we describe the experimental colloidal system. Section \ref{secResults} presents our investigations of polycrystal structure. Firstly impurities are characterised and compared to non-impurity particles in Section \ref{sec:impurities}. Subsequently, structural modifications in the vicinity of a single impurity are considered in Section \ref{secRecovery} and the dependence of global structure upon impurity concentration is investigated in Section \ref{secGlobalstructure}. Section \ref{secGrainsize} is concerned with the identification and characterisation of individual crystalline grains and Section \ref{secEntropy} presents entropy measurements based on the distribution of Voronoi cell areas in polycrystalline samples compared to that of the perfect single crystal. Finally the various lengthscales extracted from our analyses are compared and their dependence on the impurity concentration is discussed in Section \ref{secDiscussion} and conclusions are drawn in Section \ref{secConclusions}.

\section{Methods}
\label{secMethods}
The experimental colloidal sample consists of silica spheres of diameter $\sigma = 3.1 \; \mathrm{\mu m}$  and polydispersity  $s = 0.025$ obtained from Bangs Laboratory suspended in deionised water. Previous work has shown that the addition of salt to an aqueous suspension of silica colloids has little effect on the interparticle potential, and thus we neglect electrostatic interactions \cite{cui2002}. The gravitational length of this system is $l_g / \sigma = 0.005$ resulting in fast sedimentation of the particles and, at sufficient dilution, the formation of a quasi-two-dimensional monolayer adjacent to the glass wall of the microscopy cell. Once this layer is formed, out of plane motion is negligible and therefore we consider our experimental particles as a good model for hard discs. Quasi-two-dimensional colloidal samples such as this are a common model system for the investigation of two-dimensional phenomena and have previously been employed in \textit{e.g.} revealing the nature of two-dimensional phase transitions \cite{marcus1997,han2008}, confirming theoretical predictions of hard disc structure in binary mixtures \cite{thorneywork2014} and measuring the equation of state of the hard disc fluid \cite{brunner2003}.

The sample cell is constructed using glass microscope slides and coverslips held together with epoxy glue and has approximate dimensions $200 \; \mathrm{\mu m}  \times  1 \; \mathrm{cm} \times 1 \; \mathrm{cm}$. To prevent the adhesion of particles to the glass substrate against which they sediment, slides are treated with Gelest Glassclad 18 rendering them hydrophobic \cite{williams2013}. The particle stock is diluted such that the resulting monolayer is of sufficient density for the formation of an hexagonal crystal. Experimental samples are prepared at area fraction $\phi = 0.85 \pm 0.02$ where the uncertainty arises from small variations in the number of particles in the field of view and the inherent challenges in determining absolute packing fractions in colloidal experiments \cite{poon2012}.  Typically, the field of view contains $N \approx 2500$ particles. These area fractions are far in excess of $\phi=0.72$, the area fraction above which monodisperse hard discs exist as a hexagonal crystal \cite{bernard2011}, and are below the area fraction of the close packed hexagonal crystal, $\phi=\pi / \sqrt{12} \approx 0.907$.  

It is observed that a small number of the silica particles are misshapen [see Figure \ref{figPictures} (a) and detail in (b)] and thus serve as impurities, distorting the crystal structure in their vicinity. It is likely that these aspherical particles are fused dimers of the majority spherical colloids formed during the particle synthesis. A typical sample contains $< 5 \%$ impurity particles. After at least one hour of equilibration time, experimental samples are observed via brightfield microscopy for $5$ minutes. All data are acquired within two hours of sample preparation. On these timescales no ageing or coarsening of the polycrystal structure is observed. Particle trajectories are subsequently extracted using particle tracking routines adapted from those of Crocker and Grier \cite{crocker1996}. Impurities are identified by considering the eccentricity of the particle images defined for a single particle as:
\begin{equation}
\label{eq:ecc}
e = \frac{\left[ \left( \sum_i m_i \cos 2 \theta_i  \right)^2 + \left( \sum_i m_i \sin 2 \theta_i \right)^2 \right]^{1/2}}{\sum_i m_i }
\end{equation}
where $i$ labels individual pixels in the acquired micrographs, $m_i$ is the brightness of pixel $i$ and the summations run over all pixels within a circular region of radius $0.65 \sigma$ of an identified particle centre. The angle $\theta_i$ is the angle between the vector joining pixel $i$ to the particle centre and an arbitrary reference axis, taken here as the positive $x$ axis.
Perfectly spherical particles have eccentricity $e = 0$ while misshapen particles have a larger eccentricity. We apply a cut-off at $e=0.3$ and consider any objects of greater eccentricity to be impurities. We offer further characterisation of the observed impurities in Section \ref{sec:impurities}. The concentration impurities is quantified using the impurity number density, $\rho_{\mathrm{i}}$, from which the lengthscale $\rho_{\mathrm{i}}^{-1/2}$ is defined. This is the mean distance between impurities and is employed as a characteristic lengthscale for the system. In addition to colloidal experiments we perform hard disc Monte Carlo simulation of a single crystal at area fraction $\phi=0.85$ in order to compare experimentally observed structures to that of the equivalent impurity-free system.

\section{Results}
\label{secResults}

\begin{figure*}[htb]
\begin{center}
\centerline{\includegraphics[width=150mm]{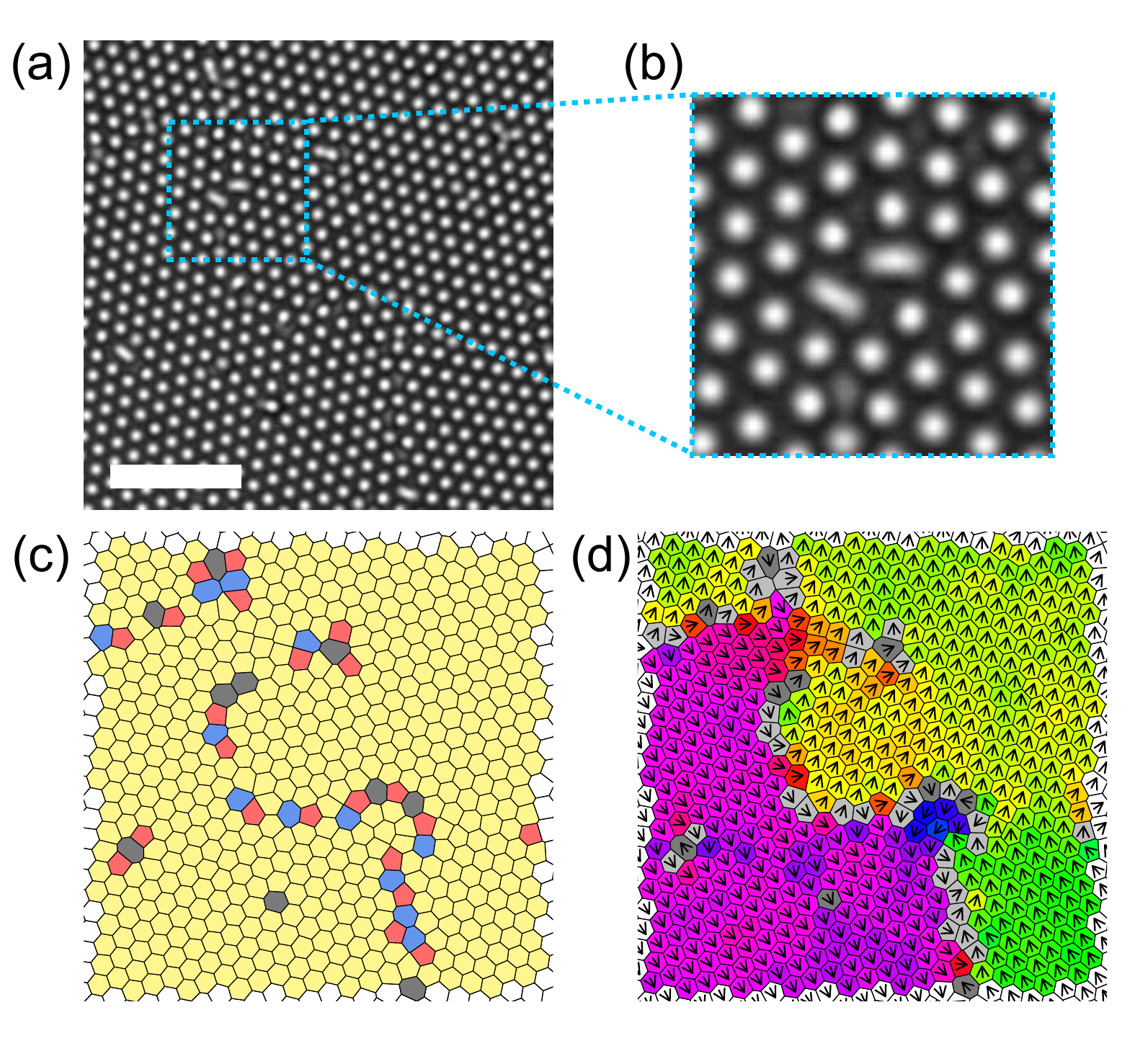}}
\caption{\label{figPictures} (Colour online) (a) Sample micrograph of polycrystalline monolayer of silica spheres. Scale bar represents $20 \; \mathrm{\mu m}$. (b) Enlarged section from (a) showing aspherical impurity particles. (c) Voronoi decomposition of the structure shown in (a). Cells are coloured according to co-ordination number (yellow: $z=6$, red: $z=5$, blue: $z=7$) except for cells corresponding to impurities (dark grey). (d) Same Voronoi tesselation as in (c) now with 6-fold co-ordinated cells coloured according to the local $\psi_6$ phase angle (also indicated by black arrows). Impurities remain dark grey while defects (5-fold and 7-fold co-ordinated cells) are light grey.}
\end{center}
\end{figure*}

At the area fractions investigated, the equilibrium structure of monodisperse hard discs in the bulk and in the absence of impurities is crystalline \cite{bernard2011,engel2013}, which holds even for quasi-two-dimensional colloids \cite{qi2014}. Instead of a single hexagonal crystal, what is observed in experiment is a polycrystalline system consisting of hexagonal domains separated by disordered grain boundaries, as shown in Figure \ref{figPictures} (a). Impurities tend to deform the hexagonal lattice in their vicinity [Figure \ref{figPictures} (a) and (b)] and are often observed at grain boundaries. We identify defects in the crystalline lattice through the local co-ordination number, $z$, found via a Voronoi decomposition of the system such as is shown in Figure \ref{figPictures} (c). Here yellow cells represent six-fold co-ordinated particles while red and blue cells are five-fold and seven-fold co-ordinated respectively and are considered to be defects. Cells corresponding to impurities are grey. It is clear from the example Voronoi tesselation shown in Figure \ref{figPictures} (c) that neighbours of impurities are often defects. This shall be elaborated upon and quantified shortly.

Local hexagonal structure within the sample is quantified using the complex bond-orientational order parameter, $\psi_6^j$, defined for particle $j$ as $\psi_{6}^{j}  = \frac{1}{z_j} \sum_{m=1}^{z_j} \exp{(i 6 \theta_m^j)} $ where $z_j$ is the co-ordination number of particle $j$, $m$ labels its neighbours and $\theta_m^j$ is the angle made between a reference axis and the bond joining particles $j$ and $m$. After summing the contributions from all neighbours of particle $j$, the resulting complex number can be written as $\psi_6^j = |\psi_6^j| e^{i\theta}$. The magnitude, $|\psi_6^j|$, quantifies the degree of hexagonal ordering about particle $j$ with perfect hexagonal ordering resulting in $| \psi_6^j | =1$. Spatial and temporal averaging of these magnitudes yields $\psi_6 = \langle | \psi_6^j | \rangle$, characterising the hexagonal nature of the entire system. The phase angle, $\theta$, of the complex $\psi_6^j$ represents the direction of the local alignment of the hexagonal lattice at the location of particle $j$. This local alignment for the sample shown in Figure \ref{figPictures} (a) is illustrated in the Voronoi diagram in (d), where both the colour of a cell and the arrow drawn at its centre indicate the phase angle of local $\psi_6^j$ and thus the local orientational direction. This rendering clearly shows the polycrystalline nature of the sample, with two domains of distinct hexagonal orientation separated by a string of impurities and defects forming a grain boundary.

\subsection{Characterising impurities}
\label{sec:impurities}

\begin{figure*}[htb]
\begin{center}
\centerline{\includegraphics[width=160mm]{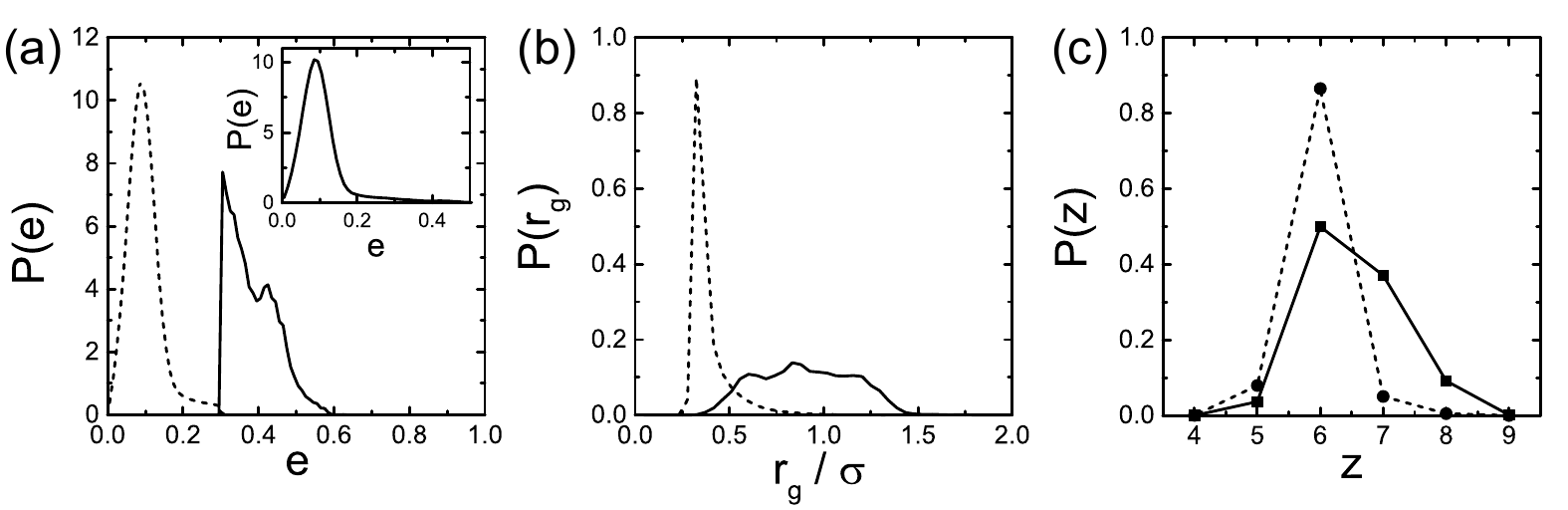}}
\caption{\label{figImpchar} (a) Particle image eccentricity distributions for impurities (solid line) and non-impurity particles (dashed line). Inset shows the eccentricity distribution for all particles, making no distinction between impurities and non-impurities. (b) Particle image radius of gyration distributions for impurities (solid line) and non-impurity particles (dashed line). Co-ordination number distribution for impurities (square points and solid line) and non-impurity particles (circular points and dashed line). }
\end{center}
\end{figure*}

Before discussing the effect of the presence of aspherical impurities on the local and global structure of the polycrystalline system it is useful to consider their size, shape and co-ordination numbers. Impurities are distinguished from the majority spherical particles by the eccentricity in their particle image, with particles having $e > 0.3$ identified as impurities. The distribution of eccentricities for all particles (impurities and non-impurities) is shown in the inset to Fig. \ref{figImpchar} (a). Only a very small number of particles have an eccentricity above the threshold. The eccentricity distributions for impurities and non-impurity particles are plotted separately in the main panel. This shows that the eccentricity of impurities, characterising deviation from a circular shape, varies between $e=0.3$ and $e \approx 0.6$ and thus all impurities are not identical in shape while the eccentricity of non-impurity particles is peaked at $e \approx 0.1$.

Figure \ref{figImpchar} (b) shows the distributions of particle image radius of gyration for impurities (solid line) and non-impurity particles (dashed line). Note that the dashed line does not peak at $\sigma / 2$ due to the fact that the particle image is composed of a bright spot surrounded by a dark halo. This measure characterises only the size of the central bright spot and is thus smaller than the true particle radius. The non-impurity distribution is sharply peaked suggesting uniformity in the size of non-impurity particles. This contrasts with the radius of gyration of impurity particle images which is much more broadly distributed, showing a wide variation in the size of impurity particles. This variation contributes to the uncertainty in the overall area fraction. The image of a particle alone is not always a good indicator of its true size, especially when the particle is irregularly shaped \cite{poon2012}, and as such estimating the area occupied by a given impurity is challenging. Related to this is the issue of quantifying the polydispersity of a population of irregularly shaped objects, which is also conceptually difficult \cite{rice2012,miller2010}. The polydispersity in a quantity $x$ is given by $s_x = \sqrt{\langle x^2 \rangle - \langle x \rangle ^2} / \langle x \rangle$ and we independently compute the polydispersities in impurity image eccentricity and radius of gyration to be $s_e = \sqrt{\langle e^2 \rangle - \langle e \rangle ^2} / \langle e \rangle = 0.20$ and $s_{r_g} = \sqrt{\langle r_g^2 \rangle - \langle r_g \rangle ^2} / \langle r_g \rangle = 0.23$. These values suggest that while there is variation in the size and shape of impurities, it is not excessive and as such we proceed in our investigation characterising each sample using only the time-averaged impurity concentration.

Figure \ref{figImpchar} (c) shows the distribution of co-ordination number for impurities (square points and solid line) and non-impurity particles (circular points and dashed line). While non-impurity particles tend to have six neighbours as expected in the hexagonal crystal, impurities have a strong tendency towards over co-ordination, that is, having $7$ or even $8$ neighbours. This is likely due to the slightly increased size of impurities over non-impurity particles and has implications for the manner in which local structure is modified in the vicinity of an impurity. This is further explored in Section \ref{secRecovery}.

\begin{figure}[htb]
\begin{center}
\centerline{\includegraphics[width=80mm]{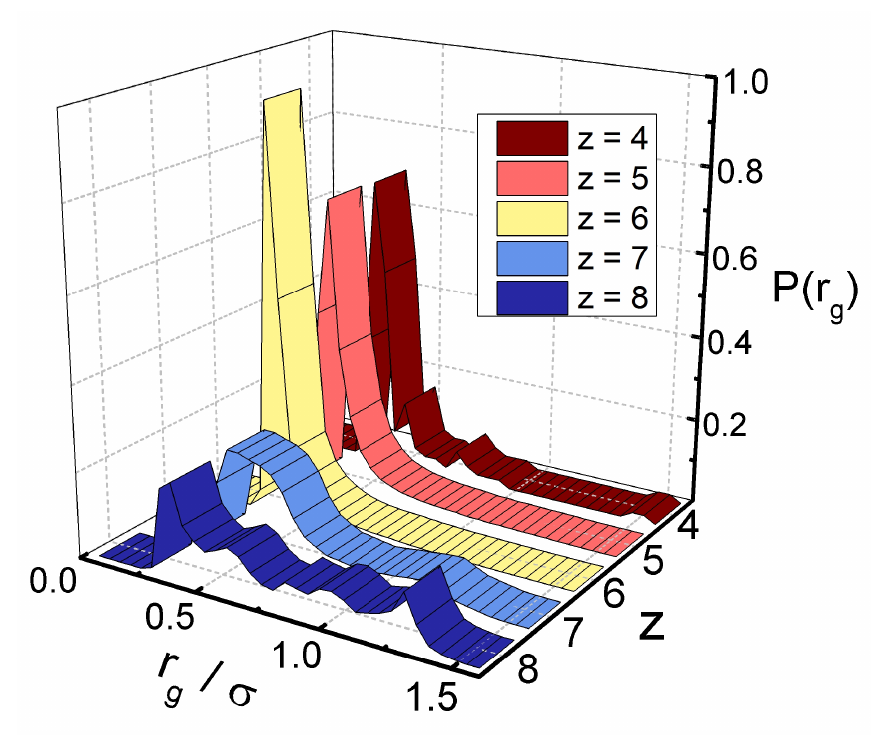}}
\caption{\label{figRghists} (Colour online) Distributions of particle image radius of gyration, $r_g$, as a function of co-ordination number, $z$, for all particle images (impurities and non-impurities). Each ribbon refers to particles having a distinct $z$ between $z=4$ and $z=8$.}
\end{center}
\end{figure}

In order to further explore the relationship between co-ordination number and particle size, Fig. \ref{figRghists} shows particle image radius of gyration distributions independently for particles having $4$, $5$, $6$, $7$ and $8$ neighbours. Here no distinction is made between impurities and the majority, spherical particles. Considering initially particles with six or fewer neighbours, a strong peak is observed at $r_g \approx 0.4 \sigma$, the same location as the peak in Fig. \ref{figImpchar} (b) for non-impurity particles. Thus it is generally the majority spherical particles that are six-fold and under co-ordinated. These distributions for over co-ordinated particles, however, show a broadening of the majority particle peak, indicating a tendency for particles with $7$ or $8$ neighbours to be larger than average --- \textit{i.e.} impurities. This finding echoes that presented in Fig. \ref{figImpchar} (c). The noise in the curves for $z=4$ and $z=8$ are due to poor statistics resulting from the rarity of these co-ordination numbers. This relationship between particle size and co-ordination number is key to understanding the modifications to local structure in the vicinity of aspherical impurities.

\subsection{Local structural recovery}
\label{secRecovery}
\begin{figure*}[htb]
\begin{center}
\centerline{\includegraphics[width=160mm]{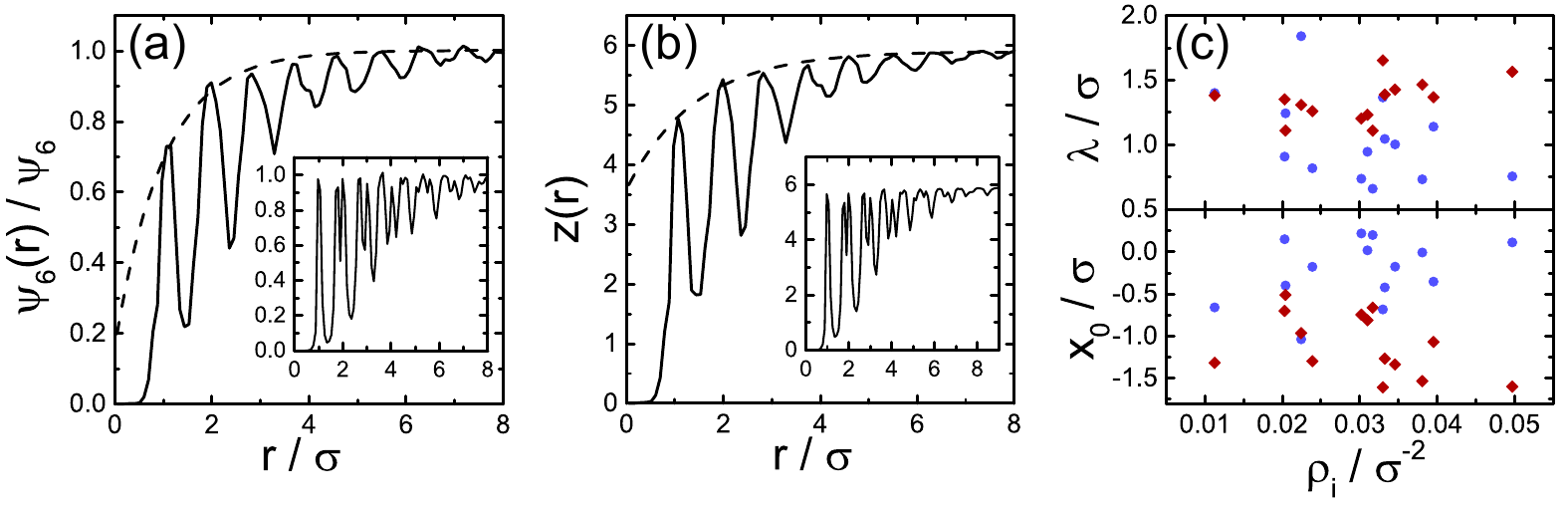}}
\caption{\label{figRecovery} (Colour online) The variation of (a) $\psi_6$ and (b) co-ordination number as a function of distance from an impurity for a single experimental sample with an overall impurity concentration $\rho_i = 0.035 \; \sigma^{-2}$. Solid line is experimental data, dashed lines are empirical fits to the peaks. Insets show the same quantities as a function of distance from a non-impurity particle. (c) Fit parameters $\lambda$ and $x_0$, defined in equation \ref{eqnRecovery}, as a function of impurity fraction as extracted from $\psi_6$ (blue) and $z$ (red). $\lambda$ is interpreted as a lengthscale for structural recovery.}
\end{center}
\end{figure*}

Firstly, we investigate structure as a function of distance from an impurity particle, essentially treating each impurity as if it were alone in the sample. Figure \ref{figRecovery} shows the variation of $\psi_6$ (a) and co-ordination number (b) as a function of distance from an impurity for a single experiment at impurity concentration $\rho_{i} = 0.035 \; \sigma^{-2}$. Both quantities are temporally and spatially averaged and $\psi_6(r)$ is normalised by the average $\psi_6$ such that it tends towards $1$ at large $r$. This approximation ignores any collective effect that multiple impurities may have on nearby particles. However, at distances less than the average distance between impurities, $r < \rho_{\mathrm{i}}^{-1/2}$, this approach is expected to be reasonable. As a comparison the insets in Figure \ref{figRecovery} show the variation in the same quantities with distance from a non-impurity particle. Both $\psi_6$ and $z$ are suppressed in the vicinity of an impurity and rise towards the average value after some distance. The suppression of the peaks in these plots is not seen in the data shown in the insets, suggesting that this effect is a direct consequence of the presence of an impurity. The implication is that, in the vicinity of an impurity, local hexagonal ordering is suppressed and average co-ordination number is reduced, indicative of defects in the hexagonal lattice. This is also evident in the Voronoi tesselation shown in Figure \ref{figPictures} (c). That these defects in the vicinity of impurities tend to be under co-ordinated (five neighbours) rather than over co-ordinated (seven neighbours) is explored below.

In order to extract a lengthscale characterising structural recovery on moving away from an impurity, the peaks of the data in Figure \ref{figRecovery} (a) and (b) are fit with a plateau function of the form
\begin{equation}
\label{eqnRecovery}
y = A\left[ 1 - \exp{\left(-\frac{(x-x_0)}{\lambda}\right)}\right]
\end{equation}
where $A$, $x_0$ and $\lambda$ are fit parameters. These fits are shown as dashed black lines in Figure \ref{figRecovery} (a) and (b). The parameters $\lambda$ and $x_0$ are shown as a function of impurity concentration in \ref{figRecovery} (c) for $\psi_6$ (blue) and $z$ (red). Neither $\lambda$ nor $x_0$ show any dependence on the impurity concentration in the range considered. Furthermore, $\lambda$ can be interpreted as a typical lengthscale for structural recovery on moving away from an impurity and is consistently of the order of $1$ particle diameter, a lengthscale smaller than the average distance between impurities, $\rho_i^{-1/2}$. 

\begin{figure}[htb]
\begin{center}
\centerline{\includegraphics[width=80mm]{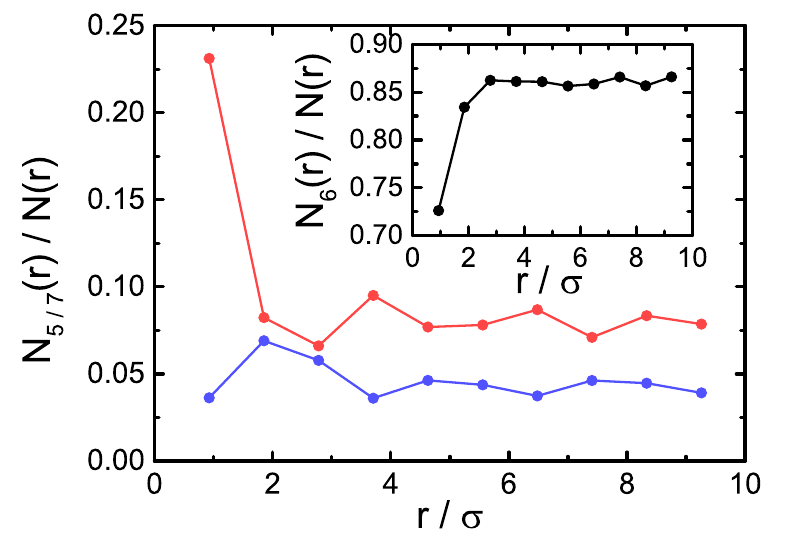}}
\caption{\label{figNeighfrominc} Average fraction of 5-fold (red) and 7-fold (blue) co-ordinated particles as a function of distance from an impurity in a single experiment at impurity concentration $\rho_i = 0.035 \; \sigma^{-2}$. Inset shows average fraction of 6-fold co-ordinated particles as a function of distance from an impurity.}
\end{center}
\end{figure}

That Figure \ref{figRecovery} (b) shows a reduction in the average co-ordination number on approaching an impurity suggests a preference for 5-fold over 7-fold co-ordinated defects directly adjacent to an impurity. Figure \ref{figNeighfrominc} plots the local average defect fractions, $N_5 / N$ and $N_7 / N$, as a function of distance from an impurity for the same experiment previously considered in Figure \ref{figRecovery} (a) and (b) and indeed supports this interpretation. For the experimental data shown, a particle that neighbours an impurity is 5-fold co-ordinated $\sim 23 \%$ of the time, while further from the impurity the average background level of $N_5/N \approx 0.09$ is recovered. Seven-fold co-ordination is slightly suppressed directly adjacent to an impurity before showing a small enhancement in the second nearest neighbour layer and subsequently returning to the background average level. The inset to Figure \ref{figNeighfrominc} shows the local average fraction of 6-fold co-ordinated particles as a function of distance from an impurity. As previously suggested by Figure \ref{figRecovery} (b), 6-fold ordering is suppressed in the vicinity of the impurity where the hexagonal structure is distorted. Qualitatively, this behaviour is observed in all experimental samples.

The excess of 5-fold defects adjacent to an impurity is understandable when one appreciates that, due to their eccentric shape, impurities are often themselves 7-, 8- or even 9-fold co-ordinated. In hexagonally ordered systems, 5-fold and 7-fold defects tend to appear in pairs (topologically, the pair is a dislocation) and as such, over co-ordinated ($z>6$) and under co-ordinated ($z<6$) particles have a tendency to cluster, or, in the case of grain boundaries, form strings \cite{gasser2010}. Since impurities are misshapen and tend to be over co-ordinated, it is expected that their neighbours are likely to be under co-ordinated. Furthermore, this observation also explains the small enhancement in the fraction of 7-fold defects around $r = 2 \sigma$ from an impurity, as the under co-ordinated particles adjacent to the impurity subsequently increase the likelihood of over co-ordination in the second nearest neighbour shell. Such chaining of defects to form grain boundaries is evident in the Voronoi decomposition shown in Figure \ref{figPictures} (c).

\subsection{Global structural ordering}
\label{secGlobalstructure}

\begin{figure}[htb]
\begin{center}
\centerline{\includegraphics[width=80mm]{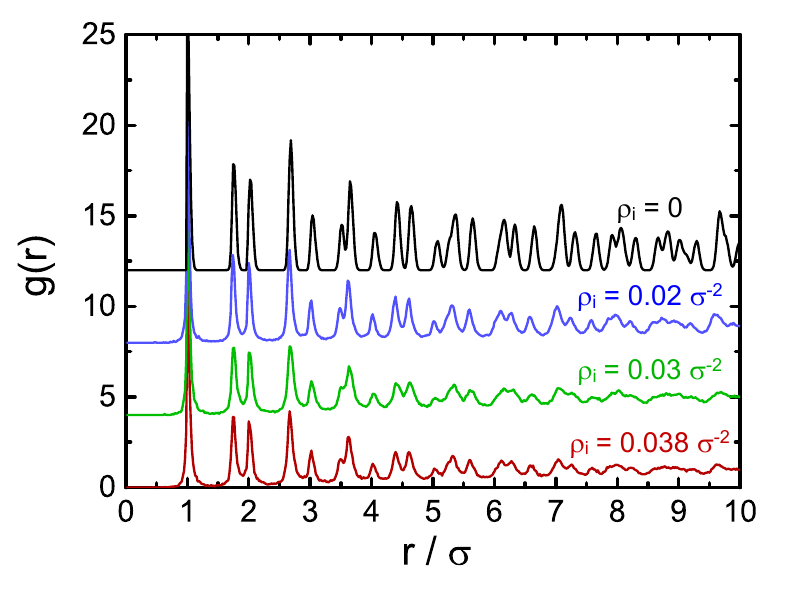}}
\caption{\label{figGr} (Colour online) Radial distribution functions, $g(r)$, for experimental samples with impurity concentration $\rho_i = 0.02 \sigma^{-2}$ (blue), $\rho_i = 0.03 \sigma^{-2}$ (green) and $\rho_i = 0.038 \sigma^{-2}$. Black line is $g(r)$ from Monte Carlo simulation of a single crystal at $\phi=0.85$. Data offset vertically for clarity.}
\end{center}
\end{figure}

Having described the effect of impurities in isolation we now consider the consequences of varying the impurity concentration on the structure of the colloidal monolayer. First we calculate the radial distribution function, $g(r)$, defined as 
\begin{equation}
g(r) = \rho^{-1} \langle \sum_{i \neq j} \delta (r_i - r_j - r) \rangle
\end{equation}
where $\rho$ is the particle number density and the indices run over all particles. This function is related to the probability of finding a particle located a distance $r$ from an arbitrary reference particle. 

Figure \ref{figGr} shows $g(r)$ for three experimental samples of differing impurity concentration (blue, green and red lines) as well as data from Monte Carlo simulation of a single hexagonal crystal at the experimental volume fraction $\phi=0.85$ (black line). The sharpness of the first peak in the experimentally measured $g(r)$ is indicative of hard-disc-like interparticle interactions while the splitting of subsequent peaks is characteristic of the organisation of particles on an hexagonal lattice. While these peaks are well defined even at large distances in the simulated single crystal, in the polycrystalline experimental samples they become increasingly poorly defined beyond $r \approx 5 \sigma$. This indicates a loss of positional correlations on such lengthscales compared to the single crystal. The experimental curves are very similar to one another, with the exception of a slight sharpening of the peaks at the lowest impurity fraction pictured (blue). This is true of all experimental samples. 

\begin{figure*}[htb]
\begin{center}
\centerline{\includegraphics[width=160mm]{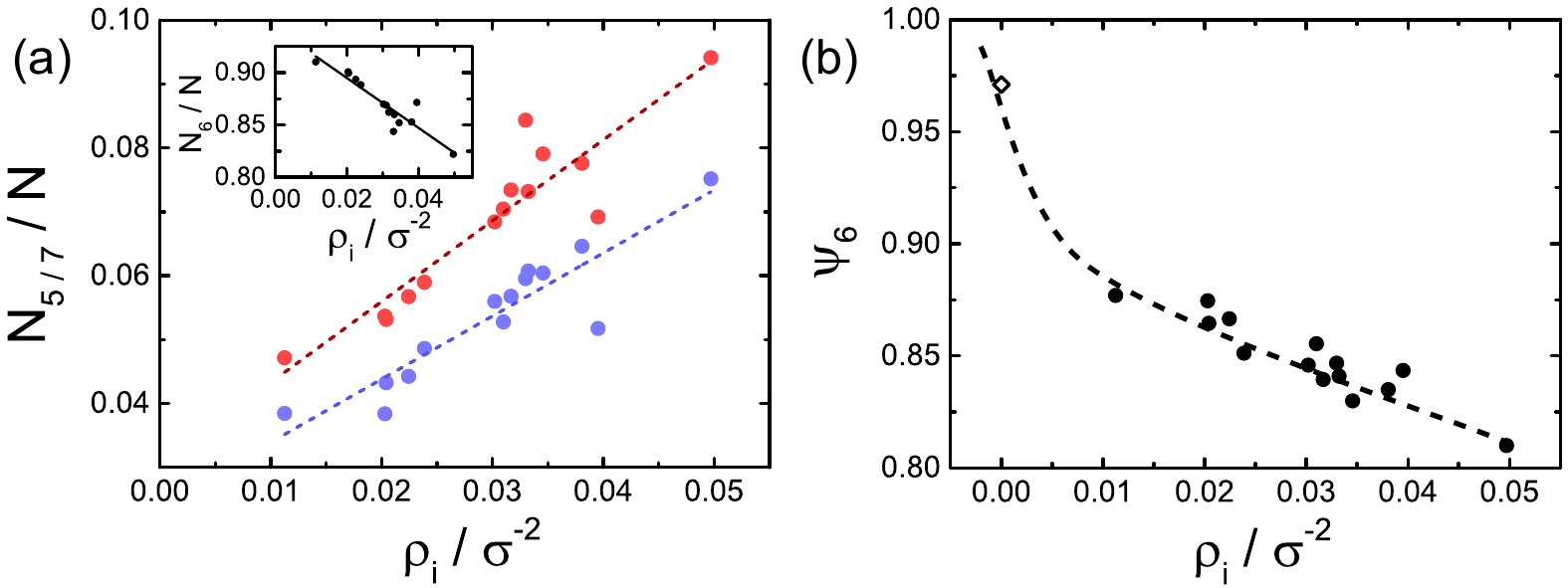}}
\caption{\label{figDefectspsi6} (Colour online) (a) Overall time-averaged fractions of 5-fold (red) and 7-fold (blue) co-ordinated particles as a function of impurity concentration. Inset shows average fraction of 6-fold co-ordinated particles. (b) The effect of impurity fraction on average $\psi_6$. Open diamond at $\rho_i= 0$ is average $\psi_6$ measured in Monte Carlo simulation of a single hexagonal crystal. Dashed line serves to guide the eye.}
\end{center}
\end{figure*}

Greater distinction between polycrystalline structures at differing impurity concentration is obtained by considering the defect populations. Figure \ref{figDefectspsi6} (a) shows the variation of the fraction 5-fold (red) and 7-fold (blue) co-ordinated particles with impurity fraction. An increase in the number of impurities leads to a similar increase in both 5- and 7-fold defects and a corresponding reduction in the fraction of particles having six neighbours (inset). In all samples there is an excess of 5-fold defects over their 7-fold counterparts, which is attributed to the preference for under co-ordination directly adjacent to an impurity.

In addition to increasing the incidence of defects in the hexagonal lattice, an increase in impurity concentration reduces the average $\psi_6$ of a sample, as shown in Figure \ref{figDefectspsi6} (b). These two observations are related to one another, as $\psi_6$ calculated for a defect particle is likely to be reduced compared to that of a 6-fold co-ordinated particle in the hexagonal lattice. Hence, introducing impurities creates defects in the lattice, the presence of which reduces the average value of $\psi_6$. This is evident when $\psi_6$ measured in polycrystalline samples is compared to that of the simulated single crystal at the same area fraction [open diamond in Figure \ref{figDefectspsi6} (b)]. The simulation is free from defects and thus exhibits very high $\psi_6$. 

\subsection{Grain size}
\label{secGrainsize}
The question now arises as to whether the average size of a crystalline grain in our polycrystalline monolayers depends on the fraction of impurities. This question is approached in two ways:  through orientational correlations in local structure and by directly identifying and subsequently characterising the polycrystalline grains.

\begin{figure*}[htb]
\begin{center}
\centerline{\includegraphics[width=160mm]{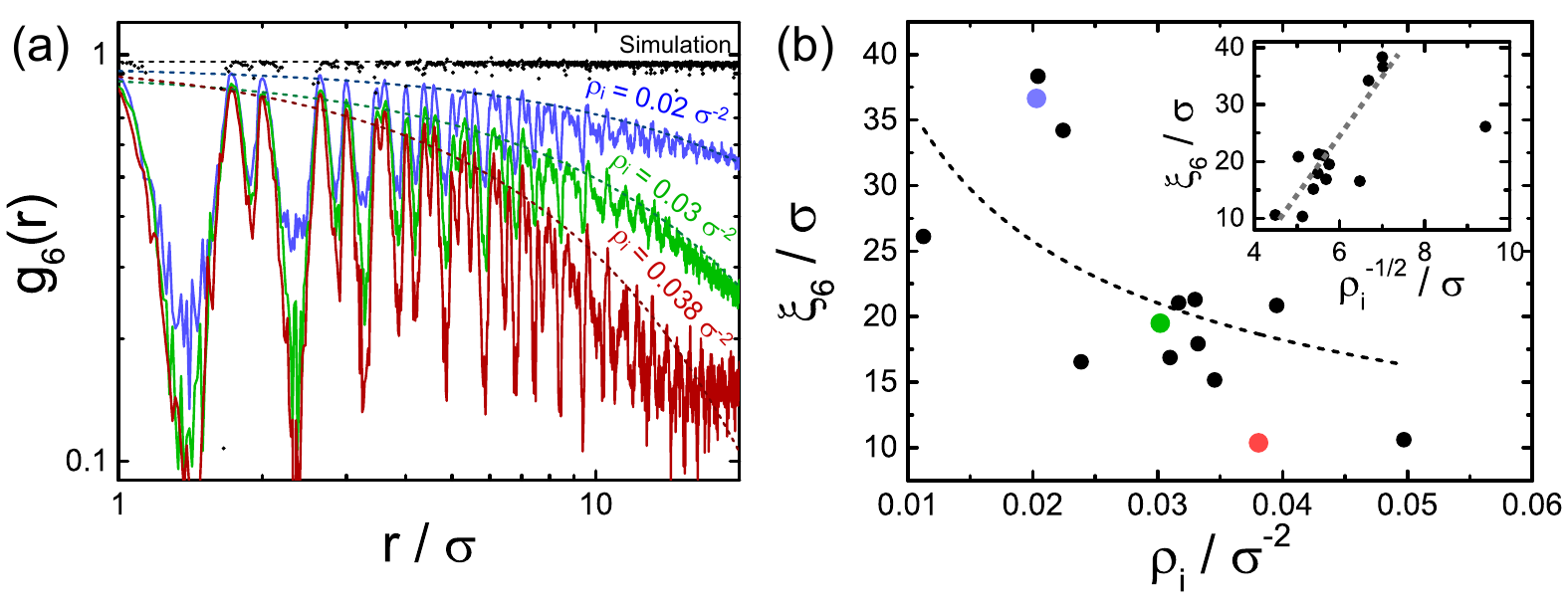}}
\caption{\label{figG6Xi6} (Colour online) (a) Orientational correlation functions, $g_6(r)$, for experimental samples with impurity concentration $\rho_i = 0.02 \sigma^{-2}$ (blue), $\rho_i = 0.03 \sigma^{-2}$ (green) and $\rho_i = 0.038 \sigma^{-2}$. Dashed lines are exponential fits to the peaks. Black points and line represent $g_6(r)$ measured in Monte Carlo simulation of a single hexagonal crystal. (b) Orientational correlation length from exponential fits to $g_6$ as a function of impurity fraction. Blue, green and red points correspond to the blue, green and red data plotted in (a). Dashed line is inverse square root fit. Inset shows $\xi_6$ as a function of the mean distance between impurities.}
\end{center}
\end{figure*}

When moving across a grain boundary, a change in the orientation of the hexagonal lattice is typically seen, as shown in Figure \ref{figPictures} (d) where the Voronoi cells are coloured based on the local phase angle of $\psi_6$. The orientational correlation function, $g_6(r) = | \langle \psi_6^j \psi_6^k \rangle |$ where particles $j$ and $k$ are separated by a distance $r$, characterises spatial correlations in this alignment and is typically used to classify the range of orientational ordering. Fluids exhibit only short-ranged orientational order and thus exponential decay of $g_6(r)$, while in the hexatic phase $g_6(r)$ decays algebraically with distance indicating quasi-long-ranged ordering. A single hexagonal crystal results in a non-decaying $g_6(r)$ \cite{gasser2010,binder2002}.

Figure \ref{figG6Xi6} (a) shows three representative orientational correlation functions calculated using experimental data and $g_6(r)$ measured in Monte Carlo simulation of a single hexagonal crystal at $\phi=0.85$ (black points). In the single crystal this function is non-decaying, indicating long ranged orientational ordering. For our polycrystalline samples at comparable area fraction, however, the envelope of the orientational correlation function is always best fit with an exponential decay [dashed lines in Figure \ref{figG6Xi6} (a)], as previously reported for quasi-two-dimensional polycrystalline colloidal systems \cite{dillmann2008}. These fits are of the form $g_6(r) \sim \exp(-r/\xi_6)$ where $\xi_6$ is the correlation length characterising a typical distance over which orientational order decorrelates. This is our first measure of polycrystalline grain size. This correlation length is plotted as a function of impurity fraction in Figure \ref{figG6Xi6} (b). A downwards trend in $\xi_6$ on increasing $\rho_i$ is evident suggesting that orientational correlations decay over a shorter range when more impurities are present indicating smaller grains. The scatter in these data is attributed to a combination of factors including small variations in area fraction between experiments and the fact that many larger grains extend beyond the finite field of view of the microscope. At low impurity concentration it is expected that $\xi_6$ will increase steeply as the polycrystal becomes increasingly monocrystalline and the orientational correlation function qualitatively changes from exponentially decaying to non-decaying. This is reflected in the dashed line in Figure \ref{figG6Xi6} (b), which varies as $\rho_{i}^{-1/2}$, relying on the assumption that the orientational correlation length is linearly related to the mean distance between impurities. The inset explicitly shows $\xi_6$ plotted against $\rho_{i}^{-1/2}$ and indeed, with the exception of a small number of outlying points, reveals an approximately linear relationship between these two lengthscales.

\begin{figure*}[htb]
\begin{center}
\centerline{\includegraphics[width=150mm]{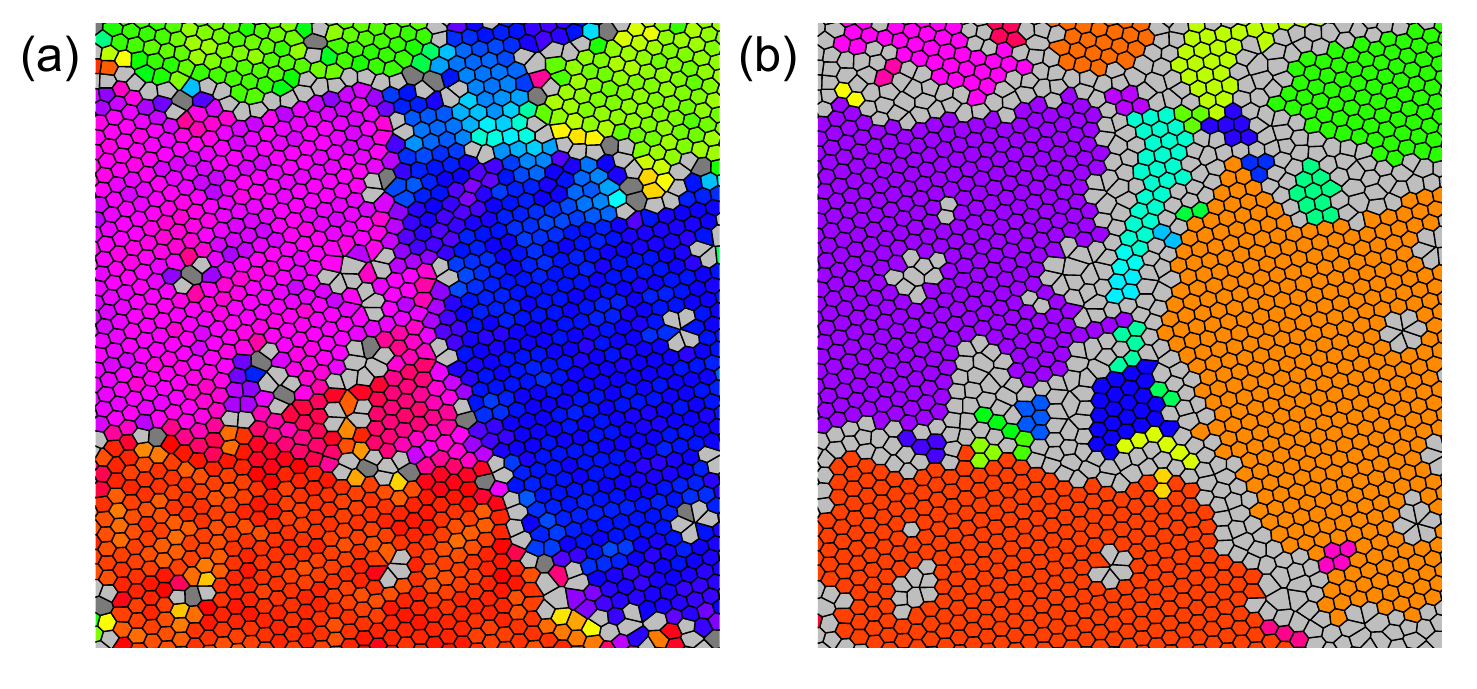}}
\caption{\label{figPolyxtal} (Colour online) (a) Voronoi tessellation of polycrystalline sample with 6-fold co-ordinated cells coloured according to the local orientation of $\psi_6$. Impurities are shown in dark grey and defects in light grey. (b) Voronoi tessellation of the same sample, now with cells coloured according to their identified polycrystalline grain. Contiguous cells of the same colour belong to the same grain. Grey cells represent particles that do not meet the criteria for inclusion in a grain.}
\end{center}
\end{figure*}

The second approach to characterising grain size requires direct identification of contiguous crystalline regions within our polycrystalline sample. Particles are identified as part of a grain if they satisfy three criteria based on those proposed by Dillmann \textit{et al.} \cite{dillmann2008,dillmann2013}. Firstly, to be considered part of a crystalline grain particle $j$ must have $| \psi_6^j | > 0.9$ and be 6-fold co-ordinated. This removes all defects and impurities from consideration and restricts the analysis to particles in strongly hexagonal local environments. The final criterion is that a particle is considered to be in the same grain as one of its neighbouring particles if the difference in the arguments of their local $\psi_6$ (the phase angle) is smaller than some cut-off. In this way, contiguous regions of particles having similar local orientational alignment are identified as single grains. The phase angle cut-off is chosen by increasing the threshold angle from $0^{\circ}$ until large grains are identified as single objects (see Figure \ref{figPolyxtal}). The chosen cut-off is $8.6^{\circ}$. It should be noted that while variations in the phase angle between neighbouring particles within a grain must be below this threshold, the local orientation within a grain may vary over long distances, as is expected in two-dimensional crystals \cite{mermin1966,mermin1968}. 

Figure \ref{figPolyxtal} (a) shows a Voronoi decomposition of an experimental sample in which the 6-fold co-ordinated cells are coloured based on the local $\psi_6$ phase angle. Regions of similar colour have similar local orientation. Defects are coloured light grey and impurities are dark grey. It is expected that contiguous regions of similar orientation should be identified as belonging to the same grain. Figure \ref{figPolyxtal} (b) shows the same Voronoi tesselation with the polygons coloured based on the crystalline grain to which they belong. Contiguous regions of a single colour represent a single identified grain. Grey cells indicate particles that do not meet the criteria for crystallinity. Identified grains can contain between $2$ and many hundreds of particles and small grains tend to be observed at the boundaries between large grains. It is expected that, given sufficient time, the polycrystalline structure may coarsen, resulting in the disappearance of these smaller grains \cite{yoshizawa2011}. However, this is not observed on the experimental timescale of five minutes.

\begin{figure}[htb]
\begin{center}
\centerline{\includegraphics[width=160mm]{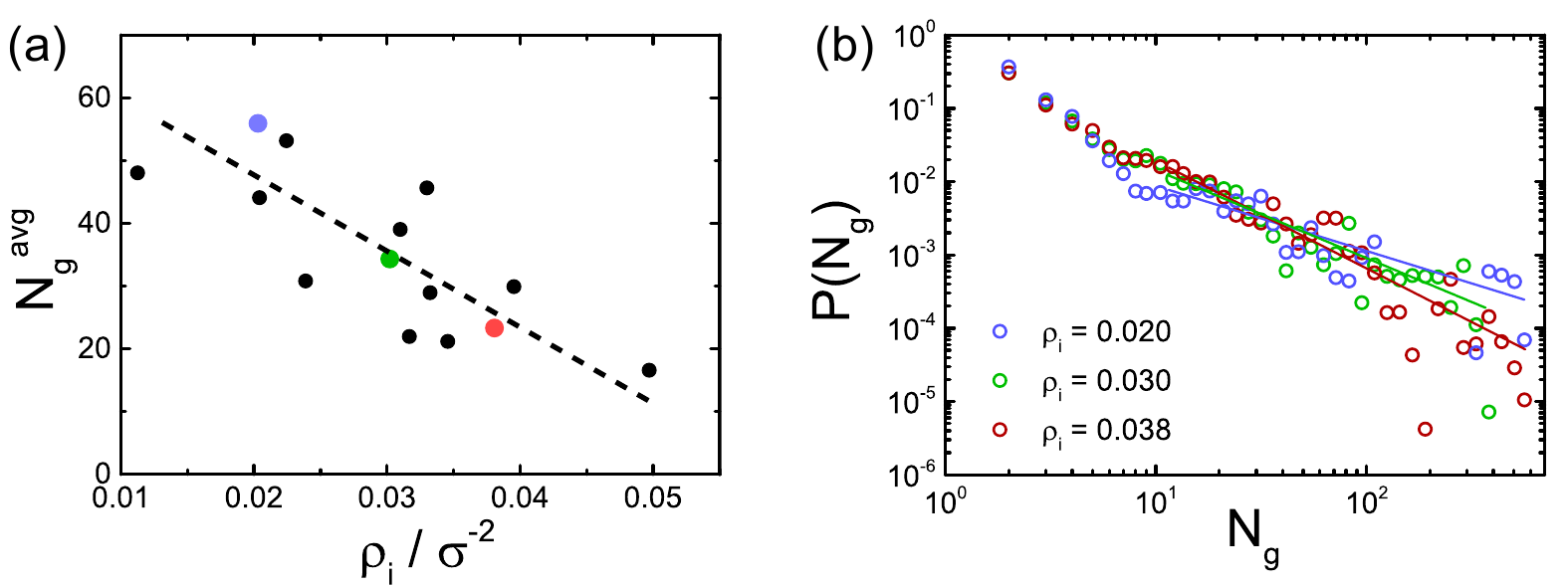}}
\caption{\label{figGrainsize}  (Colour online) (a) Average grain size, $N_g^{\mathrm{avg}}$, as a function of impurity concentration. Coloured points correspond to samples previously considered in Figures \ref{figGr} and \ref{figG6Xi6}. Dashed line indicates the downwards trend. (b) Probability distribution of measured grain sizes for these three experimental samples with impurity concentrations $\rho_i= 0.020$ (blue), $\rho_i = 0.030$ (green) and $\rho_i = 0.038$ (red).}
\end{center}
\end{figure}

Figure \ref{figGrainsize} shows the number of particles in a grain, $N_g$, averaged both over time and over all identified grains as a function of the impurity concentration. Since small `grains' consisting of very few particles far outnumber the larger grains, as is clearly visible in Figure \ref{figPolyxtal} (b), this average is much smaller than the size of the largest identified grains, which often contain many hundreds of particles. Additionally, due to the finite field of view it is expected that the true size of the largest observed grains is underestimated as these grains are likely to extend beyond the edges of the acquired images, reducing the average measured grain size compared to its true value. However, the downward trend is clear, with a greater concentration of impurities clearly resulting in smaller grains within our colloidal polycrystals. 

The full probability distribution of grain sizes is shown in Figure \ref{figGrainsize} (b) for the three experiments previously considered in Figures \ref{figGr} and \ref{figG6Xi6} (a). It is often reported that the distribution of grain areas or volumes in polycrystalline systems has a log-normal form \cite{yoshizawa2011,kurtz1980}, however this is clearly not the case when using the above criteria and grains as small as $N_g = 2$ can be identified. Instead, these distributions are more reminiscent of the crystalline cluster size distributions reported by Wang \textit{et al.} \cite{wang2010} in which clusters are also identified at the single particle level. The smallest grains are always most frequently observed and for $N_g < 7$ the three distributions shown in Fig. \ref{figGrainsize} (b) are coincident. Beyond $N_g \approx 30$, these size distributions are dominated a small number of large grains. Such behaviour is also reported by Wang \textit{et al.} at large cluster sizes. What is pertinent to the current discussion is that the size of these largest grains depends upon the impurity concentration, with the lowest impurity concentration, $\rho_i = 0.02$ (blue data), exhibiting the largest grains, and increasing impurity concentration leading to peaks at lower $N_g$. This is reflected in the average grain sizes shown in Fig. \ref{figGrainsize} (a). At large grain sizes, the distributions suggested by the data appear to have power law form with unusually broad tails, as shown by the solid lines in Fig. \ref{figGrainsize} (b). However, one must be cautious in interpreting these data as here again the finite field of view in our experiments limits the maximum observable grain size, and results in a measured grain size that is on average less than the true grain size for all but the smallest grains. Due to this effect and the short experimental duration, little comment can be made on the form of the true, underlying grain size distribution for $N_g > 30$ which is likely modified from the power law function at large lengthscales. What can be said, however, is that the largest observed grains tend to be larger in samples containing a lower concentration of impurities and that there is a downwards trend in the average grain size with $\rho_i$. 

\subsection{Entropy}
\label{secEntropy}

\begin{figure}[htb]
\begin{center}
\centerline{\includegraphics[width=80mm]{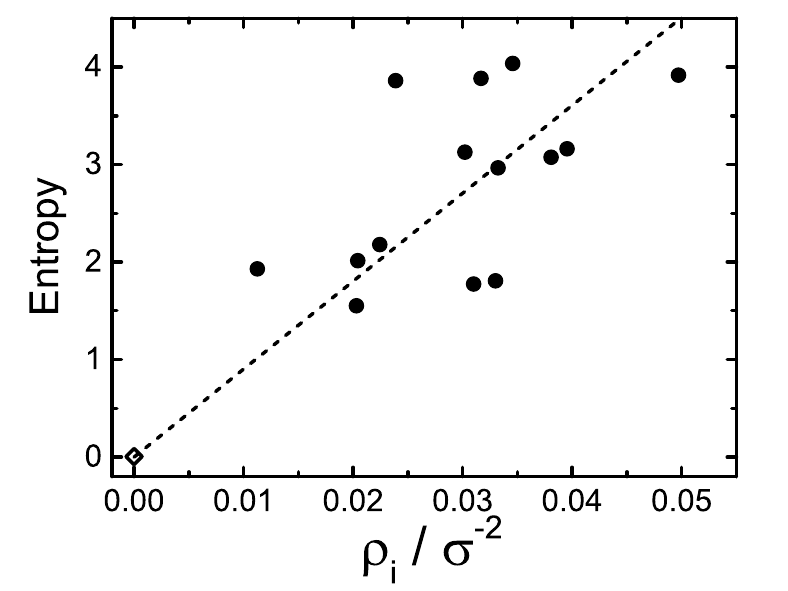}}
\caption{\label{figEntropy}  Entropy calculated from the distribution of Voronoi cell areas as a function of impurity concentration. Open diamond is calculated from Monte Carlo simulation of the a single hexagonal crystal at $\phi=0.85$ and thus is approximately zero. Dashed line is a linear fit to the data.}
\end{center}
\end{figure}

Finally we characterise the entropy of the polycrystalline system compared to that of the perfect crystal at $\phi=0.85$ based on a method developed by Aste \textit{et al.} employing a statistical mechanics approach to the distribution of single particle area fluctuations \cite{anikeenko2008,aste2008}. In this framework, the entropy is given by
\begin{equation}
S = k \left[ 1 + \ln \left( \frac{|\bar{A} - A_{\mathrm{ideal}}|}{k \Lambda^2} \right) \right]
\end{equation}
where $\bar{A}$ is the average Voronoi cell area of non-impurity particles as measured in experiment, $A_{\mathrm{ideal}}$ is the Voronoi cell area of the ideal crystal at $\phi=0.85$, equal to the area of the hexagonal crystal Voronoi cell, and $\Lambda$ is a constant analogous to the Debye length. The parameter $k$ is calculated using the standard deviation of the distribution of the measured Voronoi area distribution, $\sigma_v$:
\begin{equation}
k = \frac{(\bar{A} - A_{\mathrm{ideal}})^2}{\sigma_v^2}
\end{equation}
This measure of entropy characterises the deviation of the observed Voronoi cell structure from that of the ideal single crystal, and thus is zero for a perfect hexagonal arrangement at area fraction $\phi=0.85$. 

Figure \ref{figEntropy} shows the measured entropy as a function of impurity concentration where we have used $\Lambda = 0.025 \sigma$. The upward trend indicates that an increase in the concentration of impurities results in a corresponding increase in the system's entropy as the resulting structure deviates more strongly from that of the perfect crystal. 

\section{Discussion}
\label{secDiscussion}
\begin{figure}[htb]
\begin{center}
\centerline{\includegraphics[width=80mm]{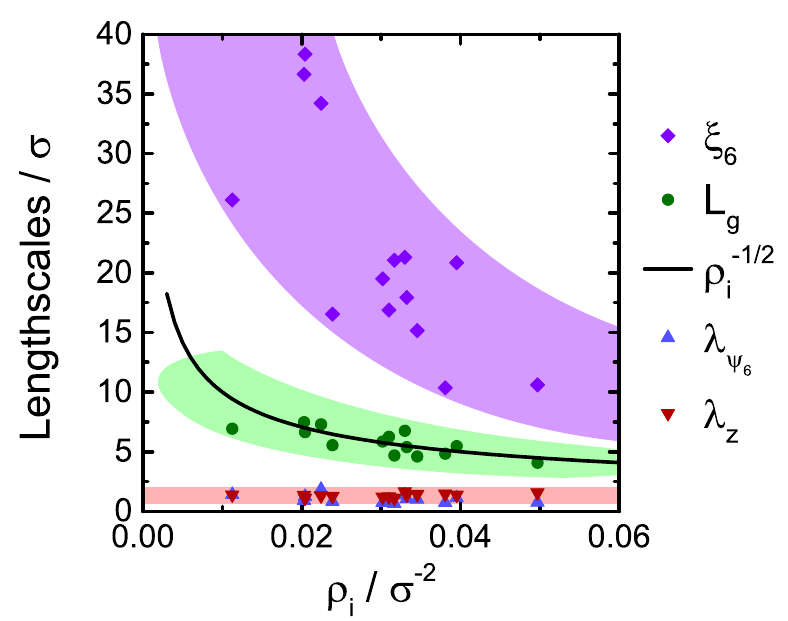}}
\caption{\label{figLengthscales}  (Colour online) Summary of the various lengthscales introduced in this research as a function of impurity concentration. Black line represents the average distance between impurities, $\rho_{i}^{-1/2}$. Purple diamonds represent $\xi_6$, previously shown in Fig. \ref{figG6Xi6} (b), purple shaded region indicates the trend and scatter in the data. Green circles are the lengthscale associated with the average grain size, $L_g / \sigma= \left(\frac{\pi}{4 \phi}\right)^{1/2} \langle N_{g} \rangle^{1/2}$, where $\langle N_g \rangle$ was previously presented in Fig. \ref{figGrainsize} (a). Green shaded region shows the trend in these data. Red and blue triangles are recovery lengths, $\lambda$, extracted from exponential fits to the envelopes of $\psi_6$ and $z$ as a function of distance from an impurity, previously shown in Fig. \ref{figRecovery} (c). Red shaded region spans the range of these recovery lengths.}
\end{center}
\end{figure}

Throughout our exploration of the role of impurity concentration on polycrystalline structure a number of lengthscales have been defined. Figure \ref{figLengthscales} summarises and compares these data as a function of impurity concentration. The black line is the mean distance between impurities, $\rho_i^{-1/2}$, while the coloured points show data previously presented in Figs. \ref{figRecovery}, \ref{figG6Xi6} and \ref{figGrainsize}. 

Firstly, the two lengthscales characterising grain size can be compared. The purple diamonds represent the correlation length $\xi_6$. The green circles are the lengthscale associated with the mean grain size $L_g / \sigma= \left(\frac{\pi}{4 \phi}\right)^{1/2} \langle N_g \rangle^{1/2}$, where the term in brackets accounts for the area occupied by a single particle at area fraction $\phi$. For the experimental value $\phi=0.85$, this term $\left(\frac{\pi}{4 \phi}\right)^{1/2} \approx 0.96$. While both of these lengths show a decreasing trend with increasing impurity concentration, $\xi_6$ is much larger and decreases more steeply than $L_g$. Since $\xi_6$ characterises orientational correlations it is sensitive to the size of the largest grains. Conversely, $L_g$ depends on the average grain size, which is dominated by a large number of small grains and thus is smaller than $\xi_6$. It is interesting to note that, in the range of impurity concentrations investigated, $L_g$ is very similar to the mean distance between impurities, suggesting that the smaller grains form in the gaps between nearby impurities. 

The blue and red triangles in Figure \ref{figLengthscales} represent the structural recovery lengths $\lambda_{\psi_6}$ and $\lambda_z$ extracted from exponential fits to $\psi_6$ and co-ordination number as a function of distance from an impurity previously shown in Figure \ref{figRecovery} (c). These lengthscales are independent of impurity concentration in the range considered and are consistently shorter than the mean distance between impurities. 

The implication of these observations is that there is a separation of lengthscales in two-dimensional polycrystals containing a low concentration of impurities. Local structural distortions in the vicinity of an impurity occur only over short distances compared to the inter-impurity distance and are independent of the impurity concentration. Conversely, the structure on lengthscales larger than the inter-impurity distance (that is the size of the larger crystalline grains) is observed to depend strongly on the impurity concentration.

\section{Conclusions}
\label{secConclusions}
Through experimental observations of dense colloidal monolayers we have investigated the role of impurities in determining polycrystalline structure in quasi-hard-discs. Impurities in this system are of a similar size to the majority spherical particles but are aspherical such that they may be identified via their eccentricity. Such impurities tend to be over co-ordinated, having $z > 6$ nearest neighbours and thus introduce defects to the hexagonal lattice. As a result, in the immediate neighbourhood of an impurity, particles tend to be under co-ordinated, giving an excess of 5-fold co-ordinated defects in the impurity's vicinity. We have shown that, on moving away from an impurity, the average co-ordination number and the local hexagonal bond orientational order parameter, $|\psi_6^j|$, recover to their background values after a few particle diameters.

In the range of impurity concentrations investigated we observe an increase in the number of defects on increasing the number of impurities. This results in a decrease in average $\psi_6$ as the hexagonal lattice becomes increasingly disordered and polycrystalline. Orientational correlations in such polycrystalline systems are found to decay exponentially as lattice alignment decorrelates on crossing a grain boundary. Furthermore, as the impurity concentration is increased we find the average polycrystalline grain size decreases and the system entropy increases. Although our global structural measures are limited due to the finite field of view, the trends are clear. 

Our data show that local structural modifications due to the presence of impurities are independent of their concentration, while structure on longer lengthscales (\textit{i.e} the size of polycrystalline grains) is determined by the impurity concentration. The lengthscale of this crossover is approximately the mean distance between impurities. The results of this investigation are primarily geometric in origin, arising from lattice distortions in the vicinity of misshapen impurities embedded in an hexagonal lattice of spherical particles. As such, the trends identified should be generally applicable to two-dimensional systems.

\ack
CPR and IW gratefully acknowledge the Royal Society and the European Research Council (ERC Consolidator grant NANOPRS, Project No. 617266) for financial support. Additionally IW was supported by the Engineering and Physics Sciences Research Council (UK) (EPSRC).

\end{document}